\documentclass[12pt]{article}
\usepackage{times}

\usepackage{eurosym}
 
\begin{document}

\begin{center}
\Large{ {\bf BOOK REVIEW}}
\end{center}

\vspace{0.3cm}

\noindent
{\bf Thermal convection: patterns, evolution and stability}, 
by Marcello Lappa,
John Willey \& Sons Ltd., 2010,
XIII$+$670 pp.,
\pounds150, \$215, \euro172.50, hardback
(ISBN-13: 978-0-470-69994-2), 
online
(ISBN-13: 978-0-470-74998-2), 
ebook
(ISBN-13: 978-0-470-74999-9).

\vspace{0.8cm}
Published in \textit{Geophysical \& Astrophysical Fluid Dynamics}, \\
DOI: 10.1080\/03091929.2010.506096 \\
URL: http:\/\/dx.doi.org/10.1080/03091929.2010.506096\\
Cite as: 
Simitev, Radostin D.(2010) 'Thermal convection: patterns, evolution
and stability, by M. Lappa',
Geophysical \& Astrophysical Fluid Dynamics, First published on: 16
August 2010 (iFirst)

\vspace{0.8cm}

\noindent
At the turn of the last century Henry B\'enard produced his famous
picture of the hexagonal flow pattern observed in a thin layer of
spermaceti heated from below. Ever since thermal convection, in its
various forms, has been a subject of continuously expanding research. This
interest is well-justified as thermal convection is, possibly, the
most common type of fluid flow in nature. In addition to their
numerous small scale manifestations, convective motions mix stellar
interiors and planetary atmospheres and oceans, they drive tectonic
plates and generate planetary magnetic fields. In industrial
applications, thermal convection enters wherever heat transfer is
involved. In nuclear reactors and ovens, in crystalisation processes
and casting, convection plays a crucial role.  Thermal convection is
important on a more fundamental level as well. It is one of 
the simplest and most easily controlled non-equilibrium systems, and
as such it is a testbed where many theoretical and experimental
methods for the study of pattern formation in spatially extended
systems and for the transition to turbulence have been developed.
The present monograph has as its main aim to give a comprehensive but,
at the same time, coherent account of thermal convection, starting
with two of its basic types --- convection due to temperature
dependence of density of fluids placed in a gravitational field and
convection due to temperature dependence of surface tension --- and
proceeding to include a large number of settings in which these two
basic cases may be combined.   
In my view, the author Marcello Lappa, of the Microgravity Advanced
Research and Support Center, Italy, succeeds well in presenting a
comprehensive account of the phenomenology, while his attempt to
provide a coherent theoretical framework is somewhat less convincing. 

In the Preface the author makes his approach for achieving the
aim of the book explicit: he is really concerned with creating a ``critical,
focused and comparative study of all these different types of thermal 
convection'' (p.~xvi), but he prefers to keep the mathematical
language ``to the minimum'' (p.~xvii). He is, though, well-aware of
the importance of mathematical concepts, and in chapter 1 the
governing equations of Fluid Dynamics and energy transport are derived
from first principles. Several related methods of theoretical analysis
are also briefly introduced here, namely linear and energy stability 
concepts and numerical approaches to the Navier--Stokes equations. The
chapter also includes a discussion of some concepts of nonlinear
dynamical systems, pattern formation and chaotic behaviour, as well as
of Maxwell's equations. 
In chapter 2 various aspects of gravity and surface tension, the forces
responsible for the two main types of thermal convection considered in
the book, are discussed. In connection with gravity, the
Boussinesq--Oberbeck approximation is introduced while in connection
with surface tension experiments on platforms orbiting in space are
described. The chapter proceeds with a list of several well-known
exact solutions of the Navier--Stokes equations for thermal problems,
including Hadley and Marangoni flows in horizontal layers of infinite
extent, and concludes with remarks on the effects of boundary layers. 
Chapter 3 provides several examples of thermal convection in nature
and in industrial applications reflecting the biases of the author. 
With this concludes a rather extensive coverage of the basics, and 
there is then a sequence of chapters, each devoted to the discussion
of a rather broad topic: 
the Rayleigh--B\'enard problem (Ch.\ 4), 
dynamics of thermal plumes (Ch.\ 5), 
systems heated from the side (Ch.\ 6),  
convection in inclined layers (Ch.\ 7), 
thermovibrational convection (Ch.\ 8), 
Marangoni--B\'enard convection (Ch.\ 9), 
thermocapillary convection (Ch.\ 10), 
mixed buoyancy--Marangoni convection (Ch.\ 11), 
hybrid regimes with vibrations (Ch.\ 12), 
and flow control by magnetic fields (Ch.\ 13).
In each of these chapters, the author draws on the literature and
describes in minute detail the relevant model and experimental
configurations and discusses the effects that have been discovered
as geometry, dimensions, boundary conditions, parameter values,
secondary effects, etc., are varied. In line with the declared aims of
the book, much effort has been expended to organize this empirical
knowledge into a coherent body by discussing the physical nature of
the related forces, by progressing from simpler idealized
configurations to more realistic experimental setups, and by
discussing sequences of bifurcations in the parameter space. This is
done in the hope to gain insight by finding analogies and common
physical mechanisms acting in seemingly different situations.  At the
same time, only elementary details of the corresponding mathematical
theory are given. While such comparative attempts to build a physical
theory often help the intuition, they also have well-known and widely
recognized conceptual  limitations, and the book would have benefitted
if more details of theoretical nature had been included. However, in
defence of the author, it must be acknowledged that most of the
interesting types of motion can even now only be qualitatively
described rather than be analysed in mathematical terms. Having in mind
this phenomenological approach to the topic, the most likely audience
for this monograph are scientists and engineers working on industrial
applications. However, the synthesis of such a large amount of
detail in the text is also likely to attract the interest of
theoreticians and experimentalists looking for new questions to
attack.

A number of original results are included. Especially interesting are
the numerous figures created from the author's own numerical 
simulations and specifically designed to illustrate various flow
features presented in the main text. A minor irritation is the style
of writing. While the use of English is generally good, one can find
instances where words are not fitting very well. 
The book is rather voluminous. This is partly justified, in view of
the wide range of phenomena that it covers, but some discussions 
run longer than really necessary, and 
some remarks are even redundant as numerous cross-references to other
parts of the book are made. The conscious choice to avoid mathematical
language certainly does not help to constrain the length of the text.

Since the author has been involved in problems of thermal convection in
connection with the subject of crystal growth, and in microgravity
environments these topics have received a detailed coverage. However,
the readers of \textit{Geophysical and Astrophysical Fluid Dynamics}
may be disappointed that the topic of convection in rotating flows is 
not included in the monograph. This also is a minor inconsistency in
the book, as many of the examples of thermally convecting flows
mentioned in chapter 2 are chosen from the realm of geophysics, 
convection in the Earth's core and the geodynamo problem, atmospheric
and oceanic convection and other planetary applications,  in all of
which the Coriolis force plays a crucial, even a dominant role.

One of the most attractive features of the monograph are its detailed
index of terms and its extensive list of references. Nearly 1300
articles and books are cited which allow the reader to enter the
literature in a very detailed way. However, in several instances the
author has missed the opportunity to reflect a different point of view
by citing other relevant monographs. For instance, the book by
Getling [1] provides a more theoretical
perspective to the topic discussed in chapter 4. Similarly the books
Gershuni and Liubimov [2], and P.~Colinet \textit{et al.} [3],
should be cited in relation to chapters 8 and 9, respectively.   
Also, the large list of references could become even more useful if
the page would be indicated on which the reference is cited.

In summary, while the book falls short of presenting a coherent 
theoretical framework of thermal convection, it is a treasure-trove of
phenomenological details ordered in a systematic way. It represents
the most comprehensive single-volume monograph on convection phenomena
available at the present time. I am 
glad to have the book on my shelf and I will recommend it to
anyone with interest in convection as an inspiring guide through its
myriad manifestations.

\vspace{1.0cm}
\rightline{Radostin D. Simitev}
\rightline{University of Glasgow}

\end{document}